# BIFURCATIONS OF FREE THERMAL VIBRATIONAL CONVECTION IN CYLINDRICAL FLUID LAYER IN MICRO-GRAVITY NUMERICAL AND ANALYTICAL RESEARCH

**Albert N.Sharifulin**

Department of Applied Physics
Perm State Technical University, 614000, Perm, Komsomolsky prospect, 29, Russia
E-mail:**sharifulin@pstu.ru**

## ABSTRACT

Vibration phenomena play an important role in many technical processes involving heated interfaces in the microgravity environment. The analysis of vibration effect on non-isothermal fluid in closed cavity is important for planning technological experiments in space [1]. Control and optimization of these processes critically depend on the understanding of liquid response to the vibrations.
With this aim the theoretical investigation for two kinds of problems are performed for infinite plane and cylindrical fluid layers.
We investigated simple case of the fluid response-thermal vibrational convection in a cylindrical fluid layer with rigid conducting boundaries.
It is found that steady modes of thermal vibrational convection are subjected to various bifurcations. Bifurcations cause sharp changes in heat transfer. The generalized Lorenz model is modified and used to conduct the analysis of bifurcations caused by the changing of the cavity shape and vibrational Rayleigh number. The shape of steady-state surface in space of $\{\psi, Rv, \gamma\}$ is found, where $\psi$ –is the streamfunction of mean flow, $Rv$ –is the vibrational Rayleigh number, $\gamma$ – is the cavity curvature. The solution correctly illustrates the general view of steady states surface for the parameter values corresponding to the cavity with the curvature close to zero (thin cylindrical layer). The numerical solution of the vibrational convection equations is performed for plane and cylindrical fluid layers. The results of the analysis based on the generalized Lorenz model are compared with the data obtained by direct numerical simulation. It is shown that the steady-state surface is different from that in the Lorenz model. The bifurcation curve with extremum is found. Thus, bifurcations of complex shape could be observed. This is impossible in generalized Lorenz model.

## 1. CONVECTION IN CLOSED CAVITY – THEORY OF STEADY STATES BIFURCATIONS BASED ON GENERALIZATIONS OF LORENZ MODEL

Let us consider closed cavity filled in with fluid. Assume that temperature distribution on the cavity boundaries sets steady vertical temperature gradient inside the cavity with no fluid displacement. Following Lorenz [4], the behavior of fluid can be described by the system of ordinary differential equations:

$$\begin{cases} \dot{\Psi} = -\Psi + Ra\vartheta_1 \\ \Pr \dot{\vartheta}_1 = -\vartheta_1 + \Psi - \vartheta_2\Psi, \\ \Pr \dot{\vartheta}_2 = -b\vartheta_2 + \vartheta_1\Psi \end{cases} \quad (1.1)$$

where Ra and Pr – Rayleigh and Prandtl criteria, b – positively defined geometrical parameter, $\Psi$ – stream-function, $\vartheta_1$ and $\vartheta_2$ – the characteristics of the deviation of temperature field from equilibrium. Lorenz applied this model for the study of fluid instability with high values of Ra. The chaotic behavior in the system (1) is widely known as strange attractor of Lorenz. We use (1) for the analysis of the bifurcations arising in numerical experiments applying full system of partial differential equations to free thermal and vibrational convection.

Lorenz's model (1.1) can be modified to take into consideration the variations in direction of heating [2]:

$$\begin{cases} \dot{\Psi} = -\Psi + Ra(\sin\alpha + \vartheta_1\cos\alpha - \vartheta_2\sin\alpha) \\ \Pr \dot{\vartheta}_1 = -\vartheta_1 + \Psi - \vartheta_2\Psi \\ \Pr \dot{\vartheta}_2 = -b\vartheta_2 + \vartheta_1\Psi \end{cases}, \quad (1.2)$$

In addition to the numbers of Rayleigh and Prandtl, the cavity inclination $\alpha$ ($0 \leq \alpha \leq 2\pi$) is introduced in (1.2). In spite of simplicity, this model correctly reflects the bifurcations of steady states of non-uniformly heated fluid in closed cavity for arbitrary heating directions.

System (1.2) has steady solutions with stream-functions satisfying the following equation:

$$\Psi^3 + b(1 - Ra\cos\alpha)\Psi - Rab\sin\alpha = 0 \quad (1.3)$$

The thermal components of these solutions is connected to stream-function $\Psi$ by the following relation:

$$\vartheta_1 = b\Psi/(b + \Psi^2), \vartheta_2 = \Psi^2/(b + \Psi^2).$$

For $\alpha = \pi$ (heating from the top) equation (1.3) has single solution $\Psi = 0$ corresponding to the fluid at rest. The linear analysis shows the stability of the solution for any $Ra > 0$. For high values of Rayleigh number small instabilities attenuate harmonically. Such a behavior is similar to well-known results of analysis of full equations of free thermal convection. For $\alpha = \pm\pi/2$ the single solution $\Psi = \pm(Rab)^{1/3}$ exists. For $\alpha = \pi/2$ the system (1.2) is transformed to well-studied Lorenz's model.

The stream-function $\Psi$ uniquely defines the steady solutions. Because of that, the three-dimensional phase space of the system (1.2) can be considered as one-dimensional for steady case. Equation (1.3) describes the steady-state surface in space of $(\Psi, Ra, \alpha)$. This space represents the product of phase space $\Psi$ and space of parameters $(Ra, \alpha)$. Figure 1 illustrates the shape of the steady-state surface $\Psi = \Psi(Ra, \alpha)$ built from equation (1.3) for b=8/3.

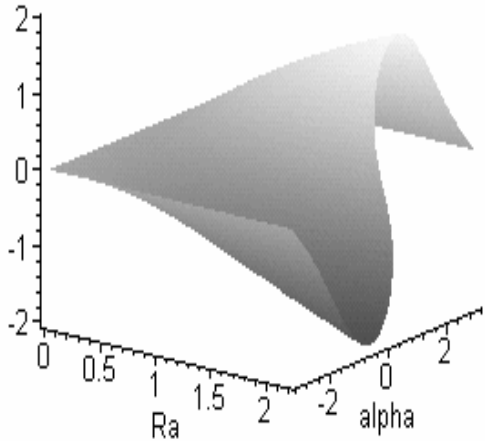

Figure 1. The shape of the steady-state surface $\Psi = \Psi(Ra, \alpha)$ for b=8/3 (built with Maple 8). The part of the steady-state surface is shown, corresponding $-\pi \leq \alpha \leq \pi$

The surface shown on Figure 1 has the Whitney cusp type of singularity for $Ra = 1, \alpha = 0$ (heating strait from the bottom). The proof based on the bifurcations theory [5] is contained in [2].

## 2. CONVECTION IN CLOSED CAVITY – BIFURCATIONS ORIGINATING FROM CHANGES IN CAVITY CURVATURE

Numerous computational experiments on system (1.2) were performed. In particular, experiments on structural stability of the system – the system response on adding to and removing out the members of the first equation. Thus, removing the component $\vartheta_2 \sin \alpha$ does not cause the qualitative change of the steady-state surface.

Let us introduce the parameter $\gamma$ and call it the cavity curvature. Assume the low and high limits for $\gamma$: -1 and +1 correspondingly. Assume no cavity curvature for $\gamma = 0$, and new system coincides with the Lorenz's model. The system for analysis of the cavity curvature can be written as

$$\begin{cases} \dot{\Psi} = -\Psi + Ra(\gamma + \vartheta_1(1-\gamma^2) - \vartheta_2\gamma) \\ \Pr \dot{\vartheta}_1 = -\vartheta_1 + \Psi - \vartheta_2\Psi \\ \Pr \dot{\vartheta}_2 = -b\vartheta_2 + \vartheta_1\Psi \end{cases} \quad , \tag{2.1}$$

The steady-state conditions of system (2.1) is derived from the equation:

$$\Psi^3 + b(1 - Ra(1-\gamma^2))\Psi - Rab\gamma = 0 \tag{2.2}.$$

The shape of the steady-state surface of the system is represented at Figure 2.

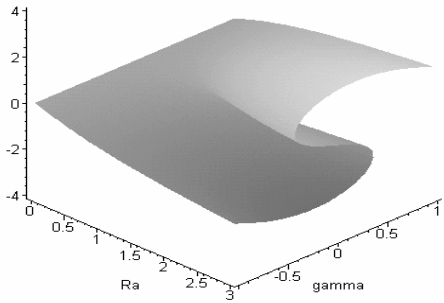

Figure 2. The shape of the steady-state surface $\Psi = \Psi(Ra, \gamma)$ for b=8/3

(built with Maple 8)

Figure 3 illustrates the projection of the steady-state surface to the parameters' plane. For the values of parameters in the light area the system (2.1) has three steady solutions. Two of them are stable, and one is unstable.

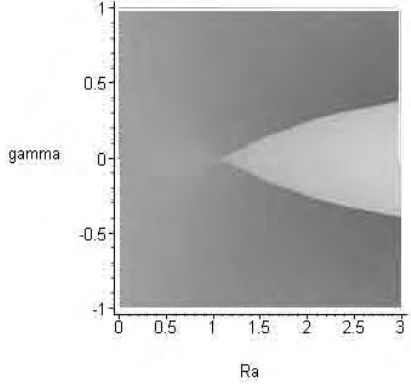

Figure 3. The areas of parameters for the system with axial symmetry

Systems (2.1) and (1.2) are coincide for small values of α and γ. We can conclude, therefore, that steady-state surface for $Ra = 1, \gamma = 0$ (heating from below and no cavity curvature) has the Whitney cusp type of singularity.

Introduced cavity curvature parameter makes the solutions of system (2.1) symmetrical $\Psi(Ra, -\gamma) = -\Psi(Ra, \gamma)$. This symmetry performs itself visually as symmetry of the steady-state surface related to the axis $\psi = 0, \gamma = 0$. The condition of symmetry can be introduced as $\Psi(Ra, -\gamma) = \Psi(Ra, \gamma)$ that means the mirror symmetry related to the plane $\gamma = 0$. For the following system of three equations

$$\begin{cases} \dot{\Psi} = -\Psi + Ra(\vartheta_1(1-\gamma^2) - \gamma^2) \\ \Pr \dot{\vartheta}_1 = -\vartheta_1 + \Psi - \vartheta_2 \Psi \\ \Pr \dot{\vartheta}_2 = -b\vartheta_2 + \vartheta_1 \Psi \end{cases}, \quad (2.3)$$

given condition is satisfied. The shape of the steady-state surface for system (2.3) is shown at Figure 5.

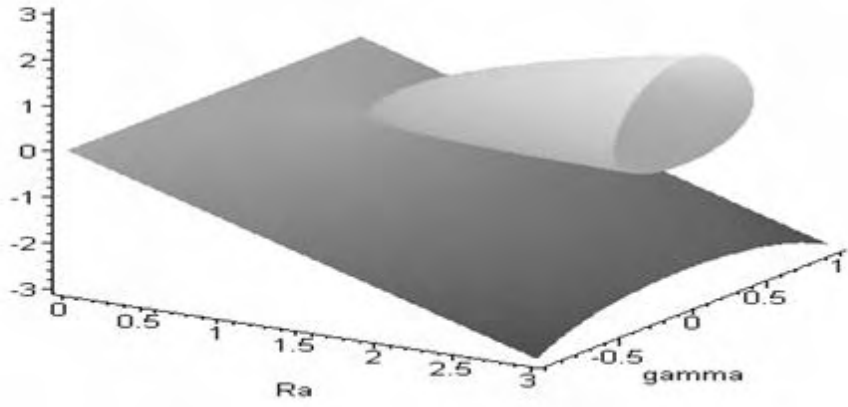

Figure 4. The steady-state surface for the system with the mirror symmetry in the area of parameters

Cross-section of steady-state surfaces, presented such Figure 4 as Figure 3 for $\gamma = 0$ are well known fork-type bifurcations diagram. By changing $\gamma$ fork type bifurcation diagram will transform in to imperfect bifurcation diagram (see Figure 5).

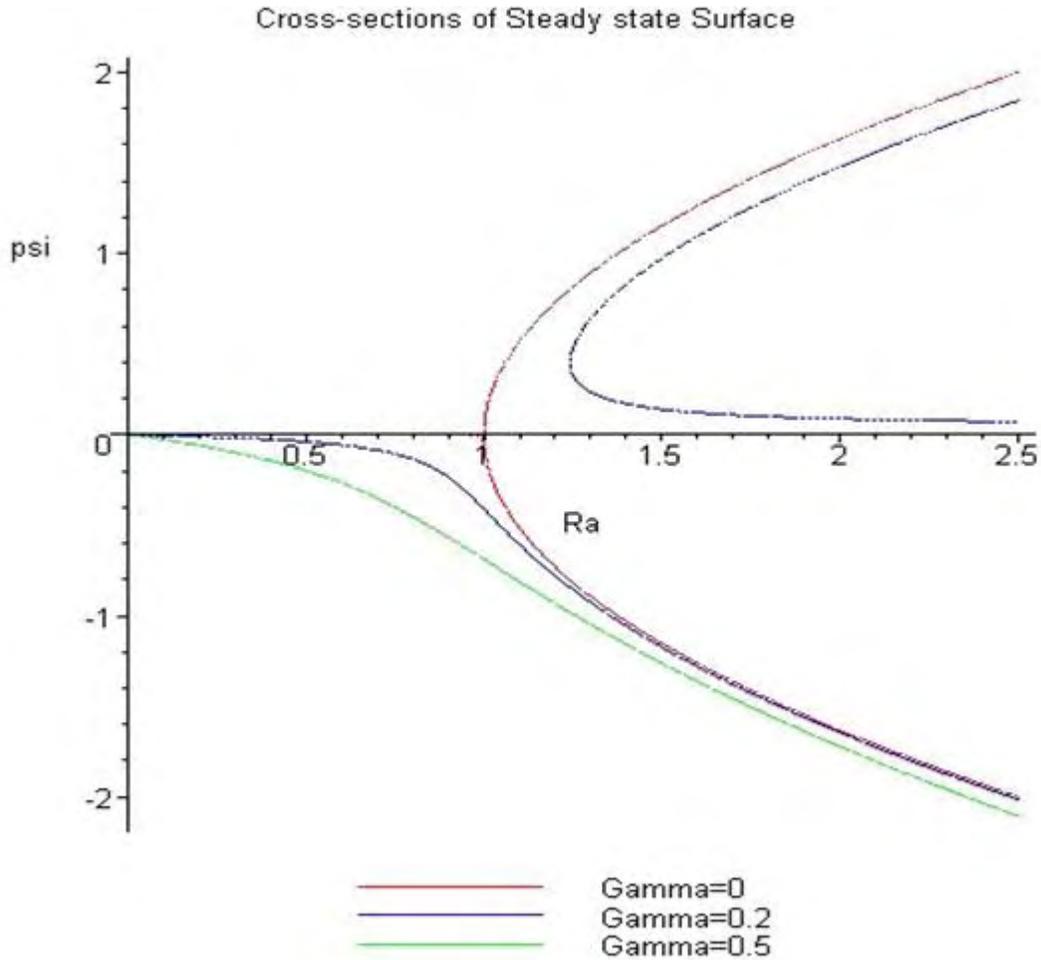

Figure 5. Perfect and imperfect bifurcation diagrams.

## 3. BIFURCATIONS OF FREE THERMAL VIBRATIONAL CONVECTION IN CYLINDRICAL FLUID LAYER IN MICRO-GRAVITY

Simple modal system based on the Lorenz's model was applied in the previous sections for simulating heat convection and bifurcations arising with smooth variations of parameters. Let us now consider bifurcations caused by the distortion of cavity using example of numerical solution for vibrational convection equations in micro-gravity for cavities representing cylindrical and plane layers.

Viscous uncompressible fluid fills in the area bounded by the infinite coaxial cylinders with radii $r_1 = a \cdot d$ and $r_2 = (a+1) \cdot d$ (cylindrical layer) or by the parallel planes $x_1 = 0, x_2 = d$ (plane layer). The boundaries of the layer are kept with constant and different temperatures $T_1$ and $T_2 = T_1 + \Theta$. The vibrations with amplitude b and angular frequency $\omega$ are performed along the direction given by the unit vector $\vec{k}$. Governing equations for thermal vibrational convection in form of generalized Gelmgolts equations are[3]:

$$\frac{\partial \vec{\varphi}}{\partial t} - \nabla \times (\vec{v} \times \vec{\varphi}) = -\nabla \times (\nabla \times \vec{\varphi}) + G_v \nabla (\vec{Wk}) \times \nabla T \qquad (3.1)$$

$$\nabla \times \nabla \times \vec{\psi} - \vec{\varphi} = 0 \qquad (3.2)$$

$$\frac{\partial T}{\partial t} + \vec{v} \cdot \nabla T = \frac{1}{\Pr} \Delta T, \qquad (3.3)$$

$$\nabla \times (\nabla \times \vec{F}) - \nabla T \times \vec{k} = 0 \qquad (3.4)$$

$$\vec{v} = \nabla \times \vec{\psi} \qquad (3.5)$$

$$\vec{W} = \nabla \times \vec{F} \qquad (3.6)$$

$\vec{\varphi}$ represents vorticity, $\vec{\psi}$ - stream function of the averaged flow, $\vec{v}$ -velocity of the averaged flow, $G_v = \frac{1}{2}(b\beta\omega\Theta d/\nu)^2$ - vibrational Grashof criteria, $\Pr = \chi/\nu$ - Prandtl number, $\vec{W}$ – pulsation velocity amplitude, $\vec{F}$ - stream function of the pulsating flow, T – dimensionless temperature, β and ν - linear expansion and kinematic viscosity coefficients. Rigid boundary conditions of following form are considered.

$$r = a: \quad T = 1, \quad \psi = F = \frac{\partial \psi}{\partial r} = 0,$$
$$r = a+1: \quad T = 0, \quad \psi = F = \frac{\partial \psi}{\partial r} = 0. \qquad (3.7)$$

Problem (3.1)-(3.7) was solved by the finite difference method for plane 2D case with neglecting dependency of all variables from $z$ coordinates. All next results has been obtained for $\Pr = 1$. Some of the results of solving this problem from [3] are represented at Figure 6 below.

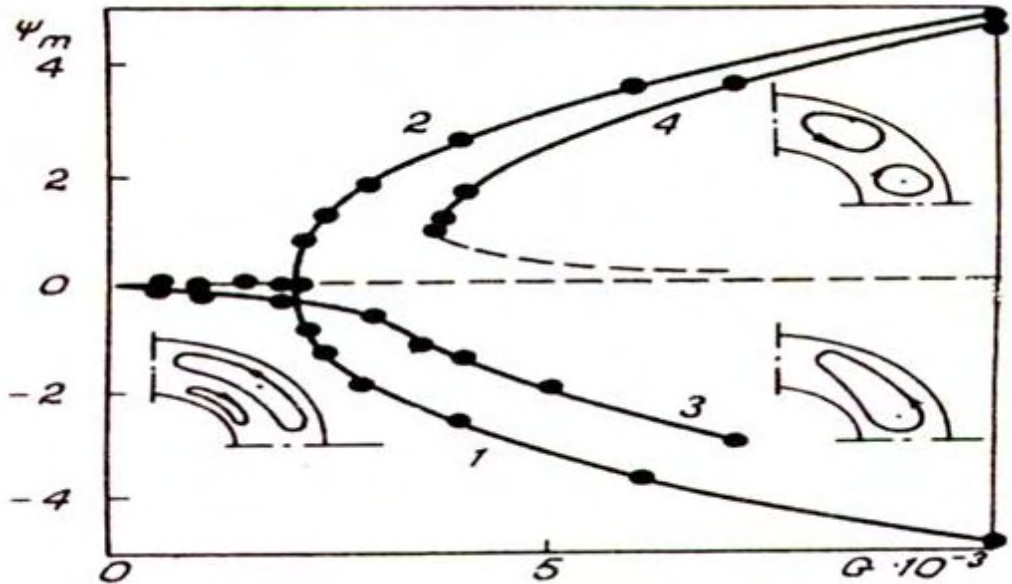

Figure 6. Stream-function maximum vs. vibrational Grashof criteria for plain (curves 1 and 2) and cylindrical (curves 3 and 4) fluid layers

It was found in numerical experiments that the structure of the over-critical averaged flow in plane layer is cellular. This flow can be observed when $G_v > 2.1*10^3$. There are two equivalent flow modes for any $G_v > 2.1*10^3$ (curves 1 and 2 at Figure 5. The mode equivalency is the result of the translation symmetry - the modes are transferred into each other with half a period translation along the direction of vibrations.

For cylindrical fluid layer with $r_2=2r_1$ the flow mode with counter-clock dominated vortex circulation is observed when $G_v < 3.6*10^3$ (curve 3). The stream function has extremum in the vortex center. There is one more stable flow mode (curve 4) with clock-wise circulation of the main vortex. Curves 3 and for are approaching curves 1 and 2 with increasing of the layer curvature $\gamma = (r_2-r_1)/(r_2+r_1)$.

The solution analysis, and in particular, analysis of $\Psi_m(\gamma)$ shows that the steady-state surface $\Psi_m(G_v, \gamma)$ has the singularity for $G_v =2.1*10^3$, $\gamma = 0$. Steady state surface have a lot of another type of singularities. This conclusion is fallowed from our current calculations presented below.

Problem (3.1)-(3.7) has been solved analytically for approach of small values of vibrational Grachof number $G_v$. In this approach the influence of mean flow on temperature field in annulus was neglected. Stream function of mean flow is given by formula:

$$\psi_1 = \frac{a^2 \sin 2\vartheta}{\ln A/a} \tilde{\psi}_1(r,a),$$

$$\tilde{\psi}_1(r,a) = \frac{A_1}{4} - \frac{A_2 s^2}{12} + \frac{A_3}{s} + A_4 s + \frac{\ln s(R + sD/\ln R)}{8D},$$

(3.8)

where $s = r^2/a^2$, $R = A^2/a^2$, $D = R - 1$,

$$A_1 = (R(4R^2 + R + 1)\ln R/D^3 - 3R(R+1)/(D \ln R) + R)/(8R),$$

$$A_3 = R^2(1/\ln R - R \ln R/D^2)/(16D^2),$$

$$A_2 = A_3 6(R+1)/R^2, \qquad A_4 = A_2/12 - A_3 - A_1/4$$

Formula (3.8) shows that along angular coordinates flows has four cells, like on Figure 7. Number of cells along radius depends from dimensionless value of inner cylinder $a$. One can obtain from (3.8) – for $a > 0.27$ (thin annulus) in annulus must be two floors of eddies. This analytical results looks like paradox. On the next figures 7 -13 presented some results of solving problem for Pr=1 by using finite-difference method on net with 20 nodes along radius and 90 – angular. Figure 7 has two floor of eddies in agreement with formula (3.8) instead $a = 1.8 > 0.27$.

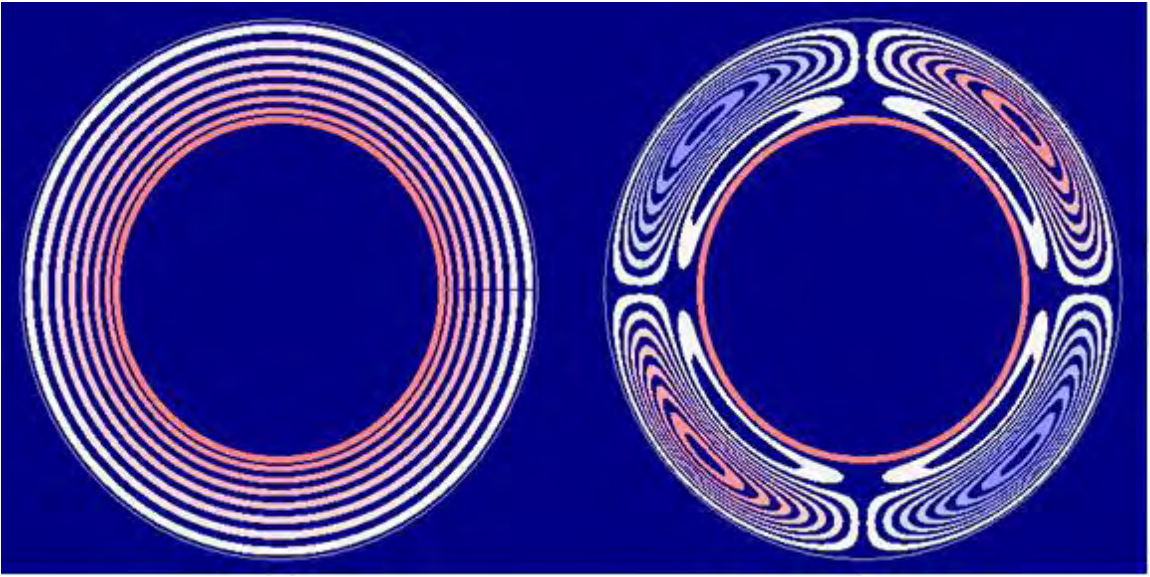

Figure 7. Isotherms and streamlines of mean flows for $G_v = 100$ and $a = 1.8$, curvature $\gamma = 0.22$. Direction of vibrations $\vec{k}$ is vertical on this and other figures.

By increasing of Grashof number $G_v$ intensity of the internal curls decreases. This structure is going to be unstable around $G_v = 3000$ and moves over to stationary state with eight cells, located in row and presented on figure 8. The most further increase of the vibrational Grashof number does not bring about change the structure of the current, but greatly redistributes heat field in annuli, see figure 9.

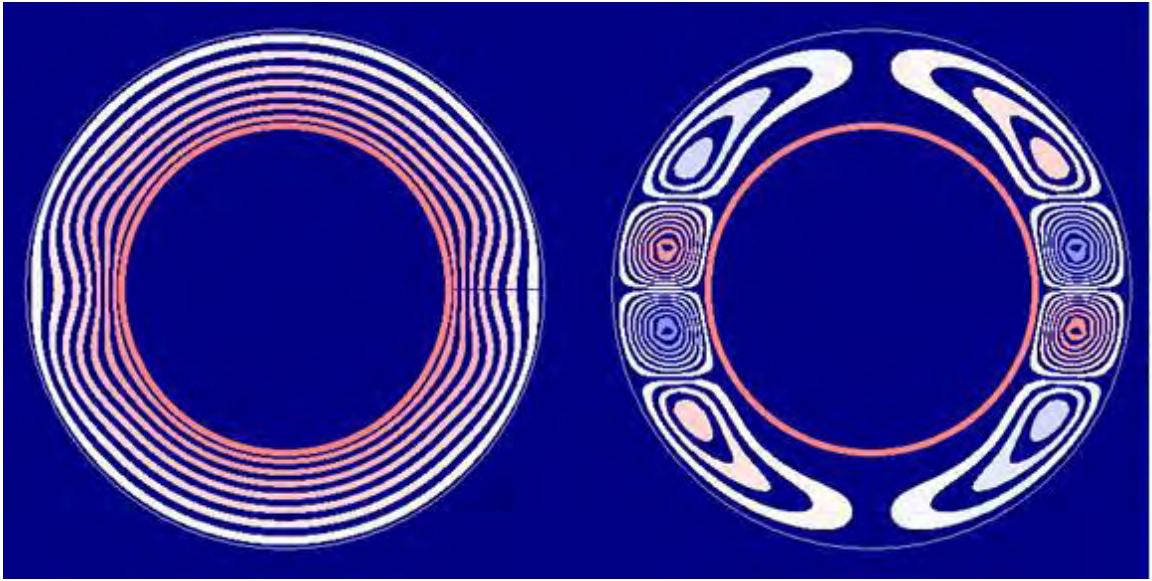

Figure 8. Isotherms and streamlines of mean flows for $G_v = 3000$ and $a = 1.8$, curvature $\gamma = 0.22$.

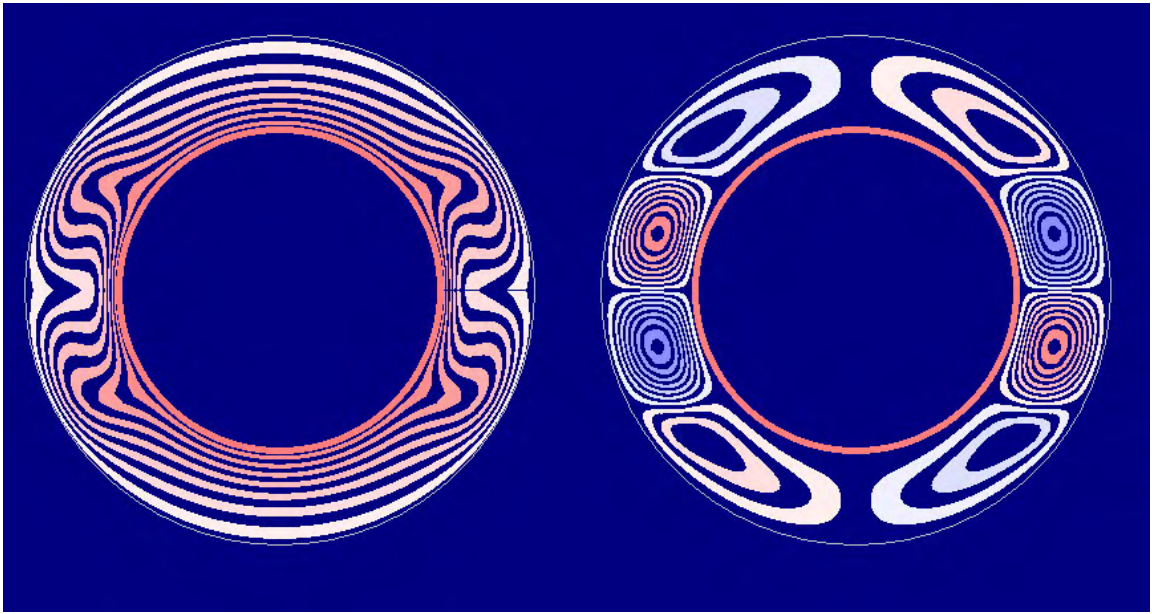

Figure 9. Isotherms and streamlines of mean flows for $G_v = 10000$ and $a = 1.8$, curvature $\gamma = 0.22$.

Results numerical experiment under fixed values of the radius of the internal cylinder $a = 1$ were submitted for figure from 10 before 13. It Is seen that under $G = 5000$ possible already three different stationary states.

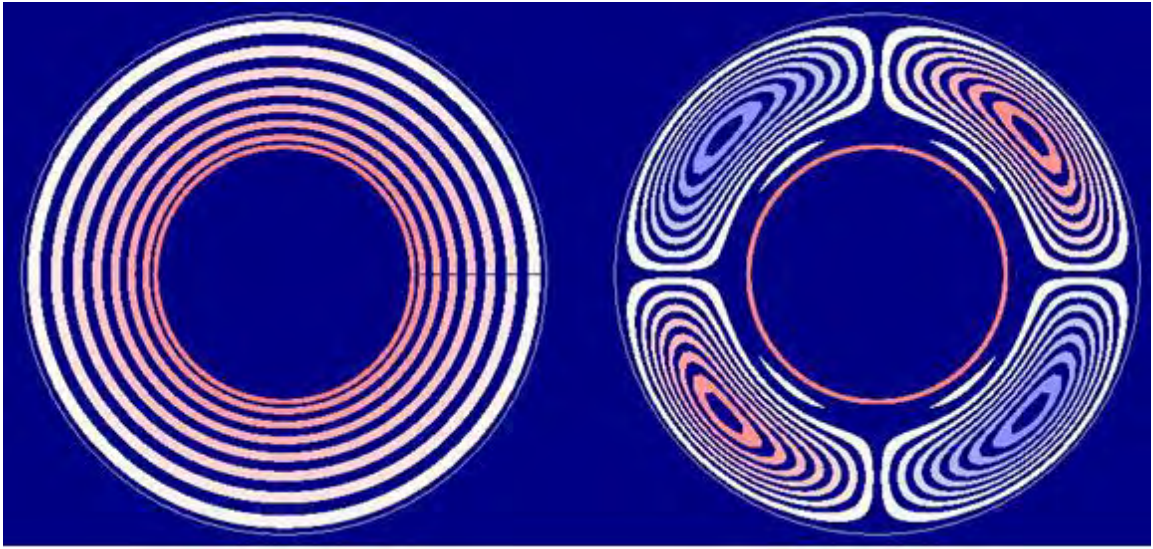

Figure 10. Isotherms and streamlines of mean flows for $G_v = 1000$ and $a = 1$, curvature $\gamma = 0.33$.

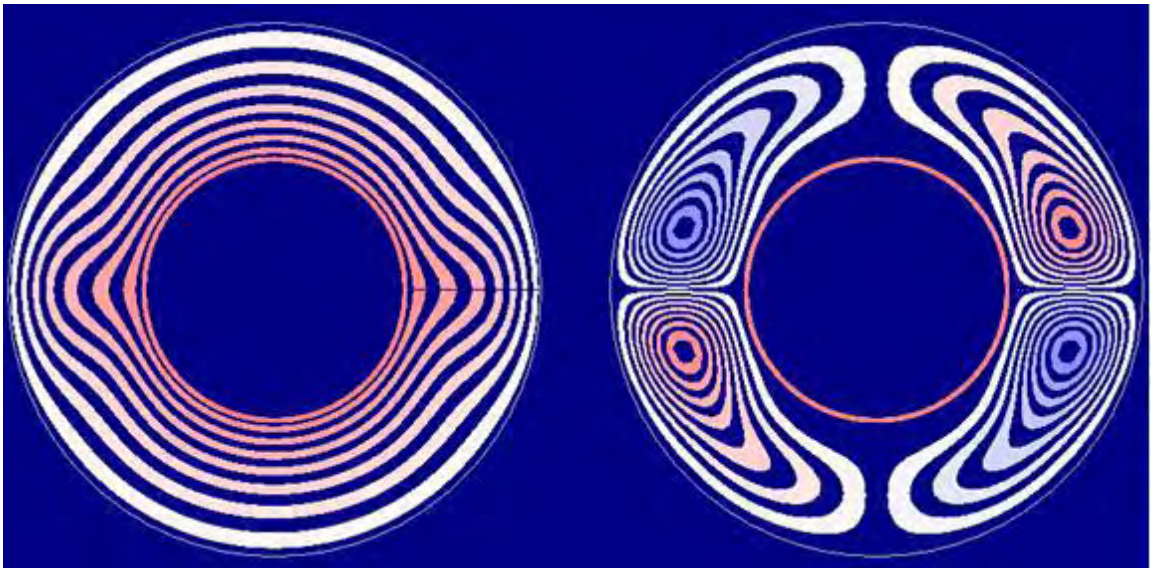

Figure 11. Isotherms and streamlines of mean flows for $G_v = 5000$ and $a = 1$, curvature $\gamma = 0.33$.

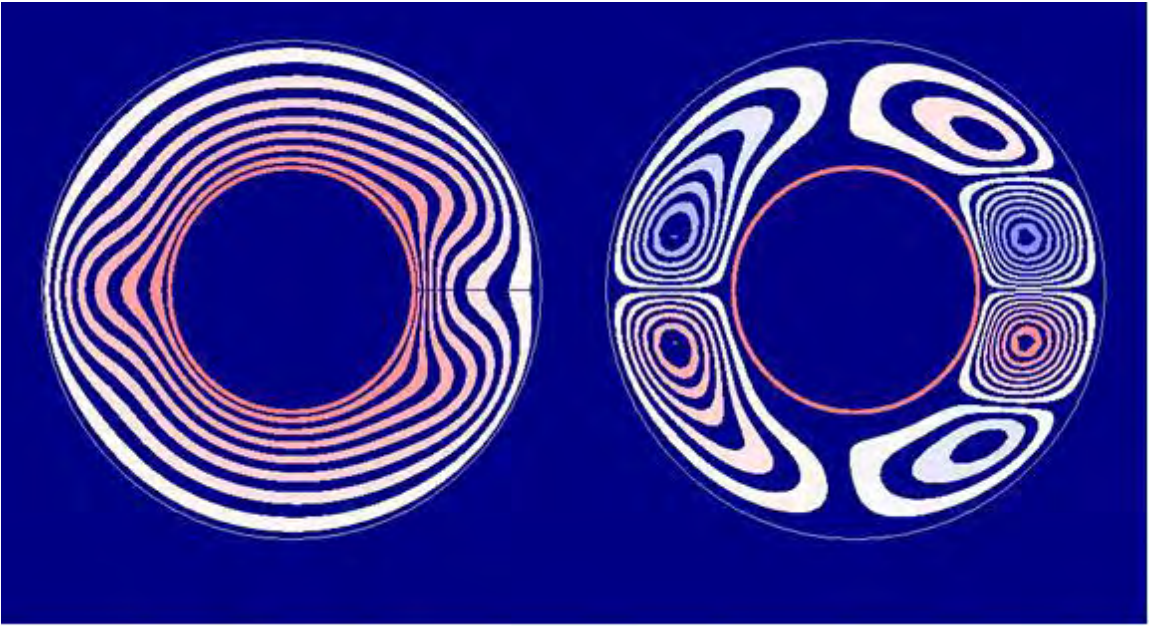

Figure 12. Isotherms and streamlines of mean flows for $G_v = 5000$ and $a = 1$, curvature $\gamma = 0.33$.

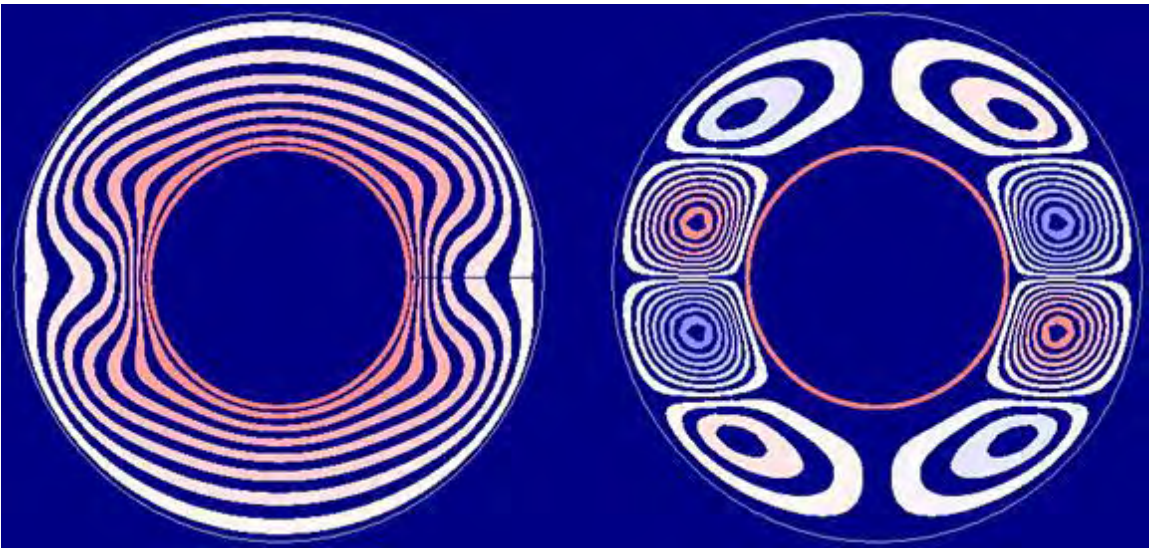

Figure 13. Isotherms and streamlines of mean flows for $G_v = 5000$ and $a = 1$, curvature $\gamma = 0.33$.

Phenomena of hysteresis was observed in computational experiments when changed curvature for fixed values of vibrational Grashof number too. On figures 14 and 15 , for example, presented steady states with and without cells for enough thick annuli for $G_v = 15000$.

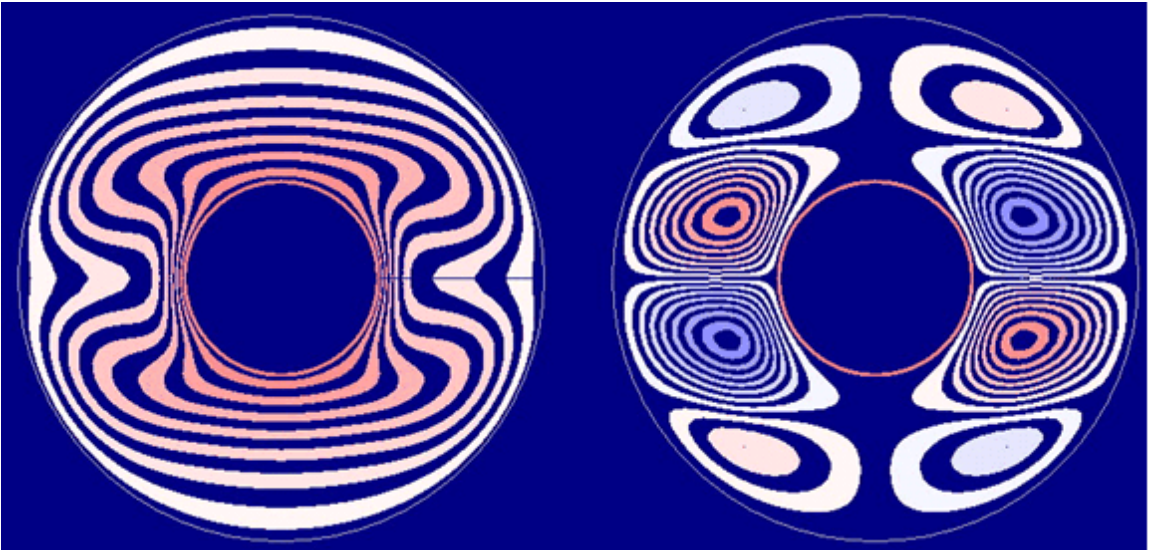

.

Figure 14. Isotherms and streamlines of mean flows for $G_v = 15000$ and $a = 0.6$, curvature $\gamma = 0.45$.

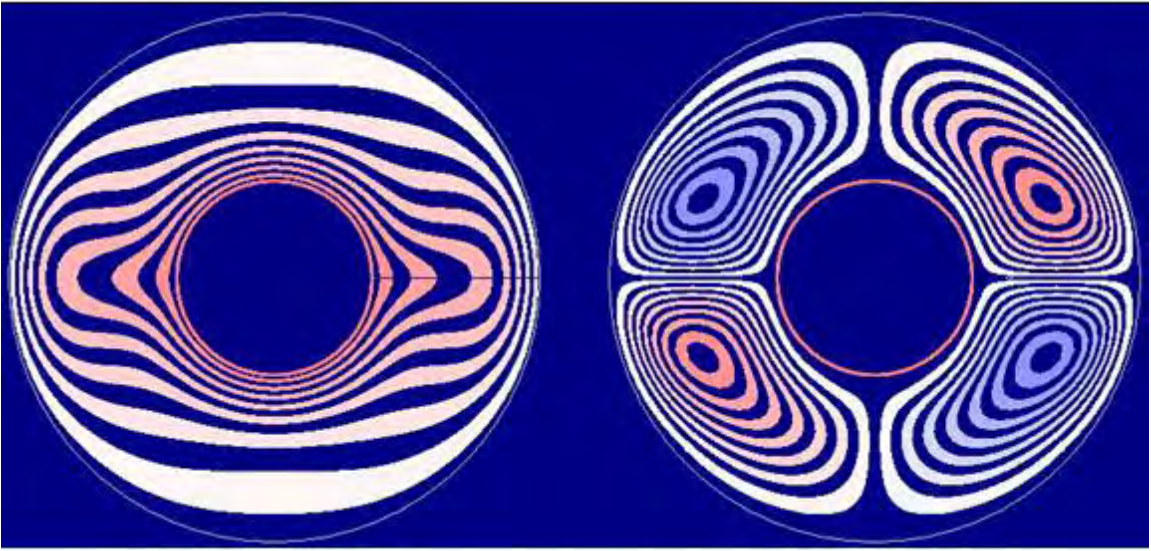

Figure 15. Isotherms and streamlines of mean flows for $G_v = 15000$ and $a = 0.6$, curvature $\gamma = 0.45$.

## CONCLUSIONS

- Zyuzgin et al.[1] had presented some results of orbital experiments (alice-vibro, Mir space station) of a near critical fluid in conditions of thermal vibrational convection. There are no possibilities for detailed comparisons, but directions of fluid flows in their experiments are in agreement with figure 10.

- The steady-state surface has much more complex shape than presented on Figure 2 and Figure 4. Phenomena of hysteresis which has been observed in computer experiments is indicative of presence of numerous additional cusps on steady-states surface of problem.